\documentclass[reprint,aps,pra,letterpaper,showpacs]{revtex4-1}
\usepackage{amssymb}
\usepackage{amsmath}
\usepackage{epsfig}
\usepackage[
pdfauthor={Paul D. Nation},
pdftitle={Nonclassical Mechanical States in an Optomechanical Micromaser Analogue},
bookmarks=true,colorlinks=true,linkcolor=blue,urlcolor=blue,citecolor=blue]{hyperref}

\begin{document}

\title{Nonclassical mechanical states in an optomechanical micromaser analogue}

\author{P. D. Nation}
\email[E-mail: ]{pnation@korea.ac.kr}
\affiliation{Department of Physics, Korea University, Seoul 136-713, Korea}
\begin{abstract}
Here we show that quantum states of a mechanical oscillator can be generated in an optomechanical analogue of the micromaser, in absence of any atom-like subsystem, thus exhibiting single-atom masing effects in a system composed solely of oscillator components.  In the regime where the single-photon coupling strength is on the order of the cavity decay rate, a cavity mode with at most a single-excitation present gives rise to sub-Poissonian oscillator limit-cycles that generate quantum features in the steady state just above the renormalized cavity resonance frequency and mechanical sidebands.  The merger of multiple stable limit-cycles markedly reduces these nonclassical signatures.   Varying the cavity-resonator coupling strength, corresponding to the micromaser pump parameter, reveals transitions for the oscillator phonon number that are the hallmark of a micromaser.  The connection to the micromaser allows for a physical understanding of how nonclassical states arise in this system, and how best to maximize these signatures for experimental observation.
\end{abstract}
\date{\today}

\maketitle

\section{Introduction}
In coupling optical and mechanical degrees of freedom via radiation pressure forces, optomechanical systems offer a promising platform by which one can prepare and control macroscopic nonclassical states of a mechanical resonator \cite{aspelmeyer:2013,poot:2012,aspelmeyer:2008}.  Experiments \cite{chan:2011,teufel:2011,oconnell:2010} have been successful in reaching the quantum ground state for a mechanical resonator using sideband cooling techniques that exploit the positive backaction of radiation pressure damping when using a pump mode that is red-detuned below the cavity resonance frequency \cite{wilson-rae:2007,marquardt:2007}.  Although superposition states have already been demonstrated \cite{oconnell:2010}, the fidelity of  these states is rapidly reduced due to unwanted heating of the mechanical resonator from the ever-present thermal environment. 

A complimentary process occurs when driving the system above the cavity resonance frequency, the blue-detuned region, where a sufficiently strong  drive gives rise to an overall negative damping rate for the mechanical oscillator leading to an instability and self-induced oscillations of the resonator that have been explored both theoretically \cite{qian:2012, nunnenkamp:2011, rodrigues:2010,ludwig:2008,vahala:2008,marquardt:2006} and experimentally \cite{anetsberger:2009,metzger:2008,aricizet:2006,carmon:2005,kippenberg:2005}.  These oscillations can be described as a mechanical, or phonon, lasing  process where a pump excitation coherently transfers energy into the mechanical oscillator leading to a threshold of oscillation, randomized phase, saturation, and a reduced linewidth \cite{khurgin:2012,vahala:2008}.  Recently, it was found that this self-oscillation regime can lead to nonclassical steady states of the mechanical resonator provided that the single-photon cavity-oscillator coupling strength is on the order of, or larger than the cavity energy decay rate and frequency of the mechanical oscillator \cite{qian:2012,rodrigues:2010}.  As steady states of the system$+$environment dynamics, these states can be measured using repeated quantum non-demolition quadrature measurements \cite{clerk:2008} and state tomography \cite{lvovsky:2009} without loss of fidelity, and could provide another route toward generating quantum states in a mechanical resonator.

Here we show that nonclassical self-oscillating mechanical states can be generated in an optomechanical system in the regime where the single-photon coupling strength is on the order of the optical cavity decay rate.  The application of sufficiently weak pump powers results in a cavity field where both the expectation value and variance of the photon number are much less than unity.  Here, photons sent into the cavity mode according to a Poisson process, using for example a laser, give rise to a  single-excitation radiation-pressure interaction that is the analogue of a micromaser \cite{walther:2006,filipowicz:1986}.  

In the micromaser, a stream of excited atoms are passed through an optical cavity one at a time according to a Poisson process generating an interaction term, $\hbar\epsilon(\hat{a}+\hat{a}^{+})\hat{\sigma}_{x}$, while the atom is inside the cavity \cite{filipowicz:1986}.  Here $\hat{a}^{+}$ and $\hat{a}$ are the bosonic creation and annihilation operators for the optical cavity mode, respectively, and $\epsilon$ is the atom-cavity coupling strength.  The emission of an excitation into the cavity mode is determined by the accumulated Rabi phase of the atom upon exiting the cavity, a quantity controlled by the product of the atom-cavity interaction time $\tau_{\rm{int}}$ and the coupling strength $\epsilon$.  In addition, the interaction time must be shorter than the inverse of the cavity decay rate; the single-excitation interaction must be quantum coherent. 

Our optomechanical analogue, consisting of two oscillator modes, relies on the fact that the use of an atom, or atom-like system, is not fundamental to operation of a micromaser.  Instead, it is the coherent single-excitation interaction that underlies the dynamics.  With the cavity mode occupied by at most a single photon, the optomechanical interaction, $\hbar g_{0}(\hat{b}+\hat{b}^{+})\hat{a}^{+}\hat{a}$, with $\hat{b}$ and $\hat{b}^{+}$ corresponding to the mechanical mode, drives the mechanical oscillator when the excitation is present, and turns off when the cavity is unoccupied, in analogy with excited atoms transiting an optical cavity.  As in the micromaser, the excitation of the mechanical oscillator is proportional to the product of the single-photon cavity-oscillator coupling strength $g_{0}$, and the effective interaction time set by the inverse of the cavity decay rate $\kappa$.  Demanding that this process be quantum coherent requires a mechanical oscillator with a large quality factor $Q_{m}$.  

Unique to the micromaser is the generation of highly nonclassical sub-Poissonian steady states of the optical cavity above-threshold when only a single excited atom is present in the cavity at any one time \cite{davidovich:1996}.  This single-atom interaction causes the steady-state photon number in the cavity to undergo a rapid increase at the onset of maser oscillations followed by a series of discontinuous jumps between stationary-states of the optical cavity with different amplitudes \cite{walther:2006}.  These jumps correspond to first-order phase transitions in the limit that the cavity damping rate goes to zero \cite{filipowicz:1986}, and are the signature of the micromaser.  We will show that all of the above features are present in our optomechanical analogue.  

Like the recent demonstration of phonon lasing in a three-mode mechanical system \cite{mahboob:2013}, this setup stands apart from the micromaser and similar systems \cite{marthaler:2011,rodrigues:2007,bennett:2006} in that there is no two- or three-level atom-like subsystem generating the maser dynamics.  Thus this work exhibits fundamental single-atom maser effects in a system composed solely of oscillator components.  Although a similar parameter regime has been considered previously \cite{qian:2012, rodrigues:2010, ludwig:2008}, the connection to the micromaser was dismissed \cite{qian:2012}. This relationship allows us to understand the onset and subsequent reduction in the nonclassical properties of the mechanical states, the role of nonlinearities, the interplay between multiple stable resonator limit-cycles, and the effect of a nonzero thermal environment on the oscillator states.  These features have not been addressed previously, and yet are important in answering the questions as to how nonclassical states arise in this system, and how best to maximize these characteristics for subsequent experimental realization.

The paper is organized as follows.  In Sec.~\ref{sec:semiclassical} we give the Langevin equations for the system, derive the nonlinear response of the cavity mode, and give expressions for the semiclassical stable limit-cycle amplitudes for the mechanical oscillator.  In Sec.~\ref{sec:quantum} we express the open quantum dynamics of the system in terms of a master equation, define an appropriate benchmark for measuring the non-classicality of the oscillator density matrix, and solve for the steady state response of the resonator as a function of detuning and coupling strength.  The results of numerical simulations are presented and results analyzed with emphasis on the micromaser analogy.  Finally, Sec.~\ref{sec:conclusion} gives a brief discussion of the results.  Details of the numerical methods used in this work are given in Appendix~\ref{sec:app}.

\section{Semiclassical Dynamics}\label{sec:semiclassical}
\begin{figure}[t]
\includegraphics[width=7.0cm]{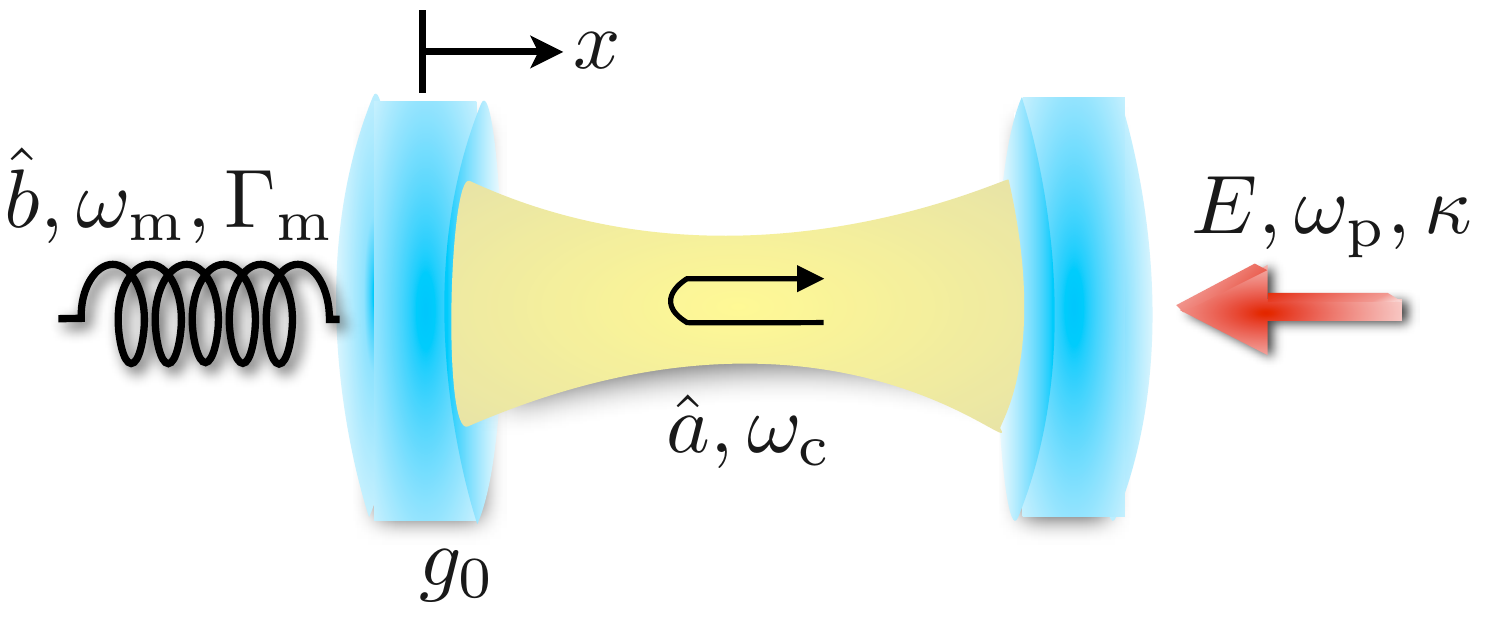}
\caption{(Color online) Conventional driven optomechanical setup where the position of a mechanical resonator ($\hat{b}$) with frequency $\omega_{m}$ and energy damping rate $\Gamma_{m}$ is coupled parametric via radiation pressure to an optical cavity ($\hat{a}$) at frequency $\omega_{c}$.  Here, the cavity is assumed to be driven by a laser with amplitude $E$ and frequency $\omega_{p}$.  The damping due to coupling with the laser mode is given by $\kappa$.}
\label{fig:fig1}
\end{figure}

Our starting point is the standard single-mode optomechanical Hamiltonian
\begin{equation}\label{eq:hamiltonian}
\hat{H}=-\Delta\hat{a}^{+}\hat{a}+\hat{b}^{+}\hat{b}+g_{0}(\hat{b}+\hat{b}^{+})\hat{a}^{+}\hat{a}+E\left(\hat{a}+\hat{a}^{+}\right),
\end{equation}
where $E$ is the pump amplitude, and we have gone into a frame rotating at the pump frequency $\omega_{p}$ and have written Eq.~(\ref{eq:hamiltonian}) in dimensionless form using the phonon energy $\hbar\omega_{m}$ where $\omega_{m}$ is the frequency of the mechanical oscillator.  The typical optomechanical setup is depicted in Fig.~\ref{fig:fig1}.  Here, all system parameters are expressed in units of the oscillator frequency.  In particular, the detuning between pump and cavity frequencies is $\Delta=\left(\omega_{p}-\omega_{c}\right)/\omega_{m}$, where $\omega_{c}$ is the resonance frequency of the cavity mode ($\omega_{c}\gg \omega_{m}$) in the linear regime.  Here, the cavity mode is assumed to be strongly-coupled to a zero-temperature environment with coupling constant $\kappa$.  In addition, we will consider a mechanical  oscillator coupled to a thermal environment with coupling strength $\Gamma_{m}=1/Q_{m}$.  Our focus will be on the  regime $g^{2}_{0}/\kappa\omega_{m}\gtrsim 1$, where the radiation-pressure of a single photon displaces the mechanical oscillator on the order of its wave packet extension \cite{aspelmeyer:2013}.  Of particular interest will be the parameter space in which the so-called granularity parameter \cite{murch:2008}, $g_{0}/\kappa\gtrsim1$, and the discreteness of the cavity photons becomes important.

Semiclassical dynamics of both the cavity and oscillator can be found using the input-output formalism \cite{gardiner:1985} to derive the Langevin equations for the operators appearing in Eq.~(\ref{eq:hamiltonian})
\begin{eqnarray}
\frac{d\hat{a}}{d\tau}&=&i\Delta\hat{a}-ig_{0}(\hat{b}+\hat{b}^{+})\hat{a}-\frac{\kappa}{2}\hat{a}- iE \label{eq:lang-cavity}\\
\frac{d\hat{b}}{d\tau}&=&-i\hat{b}-ig_{0}\hat{a}^{+}\hat{a}-\frac{\Gamma_{m}}{2}\hat{b}-\sqrt{\Gamma}\hat{b}_{\rm in}, \label{eq:lang-resonator}
\end{eqnarray}
where $\tau=\omega_{m}t$, and $\hat{b}_{\rm in}$ is the input operator for the oscillator mode.  The pump amplitude $E$ is related to the cavity input operator and input pump power via $iE=\sqrt{\kappa}\hat{a}_{\rm in}$ and $P=\hbar\omega_{p}E^{2}/\kappa$, respectively.  Setting the operators equal to their expectation values, $\bar{a}=\langle \hat{a}\rangle, \bar{b}=\langle\hat{b}\rangle$, while time derivatives and resonator input operator $\hat{b}_{\rm in}$ are set to zero, yields deterministic equations that are used to find the steady-state average photon number for the cavity mode $\bar{N}_{a}$
\begin{equation}\label{eq:nonlinear}
E^{2}=\left(\Delta^{2}+\kappa^{2}/4\right)\bar{N}_{a}-2\Delta\mathcal{K}\bar{N}_{a}^{2}+\mathcal{K}^{2}\bar{N}_{a}^{3},
\end{equation}
where we have simplified the expression using intrinsic Kerr nonlinearity of the radiation pressure coupling \cite{nation:2008}
\begin{equation}\label{eq:kerr}
\mathcal{K}=-\frac{2g_{0}^{2}}{1+\frac{\Gamma^{2}_{m}}{4}}.
\end{equation}
Here the minus sign indicates that this nonlinearity has a ``spring-softening" effect on the cavity; the cavity resonance frequency is pulled below its corresponding linear value.  Equation (\ref{eq:nonlinear}) determines the amplitude of the cavity field as a function of the detuning and gives rise to the well-known radiation pressure bistability \cite{dorsel:1983}.  One may define the renormalized cavity resonance frequency as the value for which Eq.~(\ref{eq:nonlinear}) is maximized.

For the dynamics of the system, we follow Ref.~\cite{marquardt:2006} and make the ansatz that the resonator undergoes sinusoidal oscillations obeying  $x(\tau)=\bar{x}+A\cos(\tau)$, where $\bar{x}$ is the static displacement of the resonator and $A$ is the amplitude of oscillation, both measured in units of the resonator zero-point motion $x_{\rm zp}=\sqrt{\hbar/2m\omega_{m}}$, where $m$ is the effective mass of the oscillator.  An exact solution is found using the Fourier series $\bar{a}(\tau)=e^{i\varphi(\tau)}\sum_{n=-\infty}^{\infty}\alpha_{n}e^{in\tau}$ in Eq.~(\ref{eq:lang-cavity}) with coefficients
\begin{equation}\label{eq:harmonic-cav}
\alpha_{n}=-iE\frac{J_{n}(g_{0}A)}{i\left(n-\Delta+g_{0}\bar{x}\right)+\kappa/2},
\end{equation}
and time-dependent phase $\varphi(\tau)=-g_{0}A\sin(\tau)$.  The time-averaged cavity occupation number $\overline{\langle|\bar{a}|^{2}\rangle}=\sum_{n}|\alpha_{n}|^{2}$ is therefore peaked at discrete values given by $\Delta=n+g_{0}\bar{x}$ where the integer $n$ labels the mechanical sideband, i.e. $n\omega_{m}$.  The static displacement and self-oscillation amplitudes are found by self-consistently solving the time-averaged force balance 
\begin{equation}\label{eq:force}
\bar{x}=-2g_{0}\sum_{n}|\alpha_{n}|^{2}
\end{equation}
and power balance
\begin{equation}\label{eq:power}
\Gamma_{m}A=-4g_{0}\mathrm{Im}\sum_{n}\alpha^{*}_{n+1}\alpha_{n}
\end{equation}
equations, respectively.  In general, there can be multiple stable limit-cycle amplitudes for a given set of system parameters.  For a high-Q mechanical mode, Eqs.~(\ref{eq:nonlinear}) and (\ref{eq:force}) combine to give $g_{0}\bar{x}\propto \mathcal{K}$, showing that the cavity response Eq.~(\ref{eq:harmonic-cav}) also accounts for frequency-pulling effects, although the lineshape remains Lorentzian.  

\section{Quantum Dynamics}\label{sec:quantum}
To understand the nonclassical features in the mechanical steady states of Eq.~(\ref{eq:hamiltonian}) we simulate the full quantum dynamics using the master equation
\begin{equation}\label{eq:master}
\frac{d\hat{\rho}}{d\tau}=\mathcal{L}\hat{\rho}=-i[\hat{H},\hat{\rho}]+\mathcal{L}_{\rm cav}[\hat{\rho}]+\mathcal{L}_{\rm mech}[\hat{\rho}],
\end{equation}
where the dissipative terms $\mathcal{L}_{\rm cav}$ and $\mathcal{L}_{\rm mech}$ are assumed to be in Lindblad form
\begin{eqnarray}
\mathcal{L}_{\rm cav}&=&\frac{\kappa}{2}\left(2\hat{a}\hat{\rho}\hat{a}^{+}-\hat{a}^{+}\hat{a}\hat{\rho}-\hat{\rho}\hat{a}^{+}\hat{a}\right) \\
\mathcal{L}_{\rm mech}&=&\frac{\Gamma_{m}}{2}(\bar{n}_{\rm th}+1)(2\hat{b}\hat{\rho}\hat{b}^{+}-\hat{b}^{+}\hat{b}\hat{\rho}-\hat{\rho}\hat{b}^{+}\hat{b}) \nonumber \\
&+&\frac{\Gamma_{m}}{2}\bar{n}_{\rm th}(2\hat{b}^{+}\hat{\rho}\hat{b}-\hat{b}\hat{b}^{+}\hat{\rho}-\hat{\rho}\hat{b}\hat{b}^{+}),
\end{eqnarray}
where the cavity input port is at zero temperature, and the mechanical resonator environment is parameterized by the average number of thermal excitations $\bar{n}_{\rm th}=\left[\exp(\hbar\omega_{m}/k_{\rm B}T)-1\right]^{-1}$.

As a measure for the nonclassical features in the oscillator steady-state we consider the negativity of the Wigner function \cite{haroche:2006} as measured by the ratio between the sum of negative and positive discretized Wigner densities, here called the nonclassical ratio
\begin{equation}\label{eq:ratio}
\eta=\frac{\sum_{n}|w^{(-)}_{n}|}{\sum_{m}w^{(+)}_{m}}=\frac{\sum_{n}|w^{(-)}_{n}|dxdp}{1+\sum_{n}|w^{(-)}_{n}|dxdp},
\end{equation}
where $w^{(-)}_{n}$ and $w^{(+)}_{m}$ are the amplitudes of negative and positive density components, respectively, and $dxdp$ is the area element.  The second equality in (\ref{eq:ratio}) follows from the fact that the total Wigner function must sum to one.  For the parameter regime considered here, Eq.~(\ref{eq:ratio}) is nearly linear, making it a suitable benchmark for comparing quantum signatures in the oscillator states.  To put the values of Eq.~(\ref{eq:ratio}) in context, we point out that the first Fock state $|1\rangle$ has a nonclassical ratio of $\sim 18 \%$, with all higher Fock states above this value. 

\begin{figure}[h]
\includegraphics[width=8.6cm]{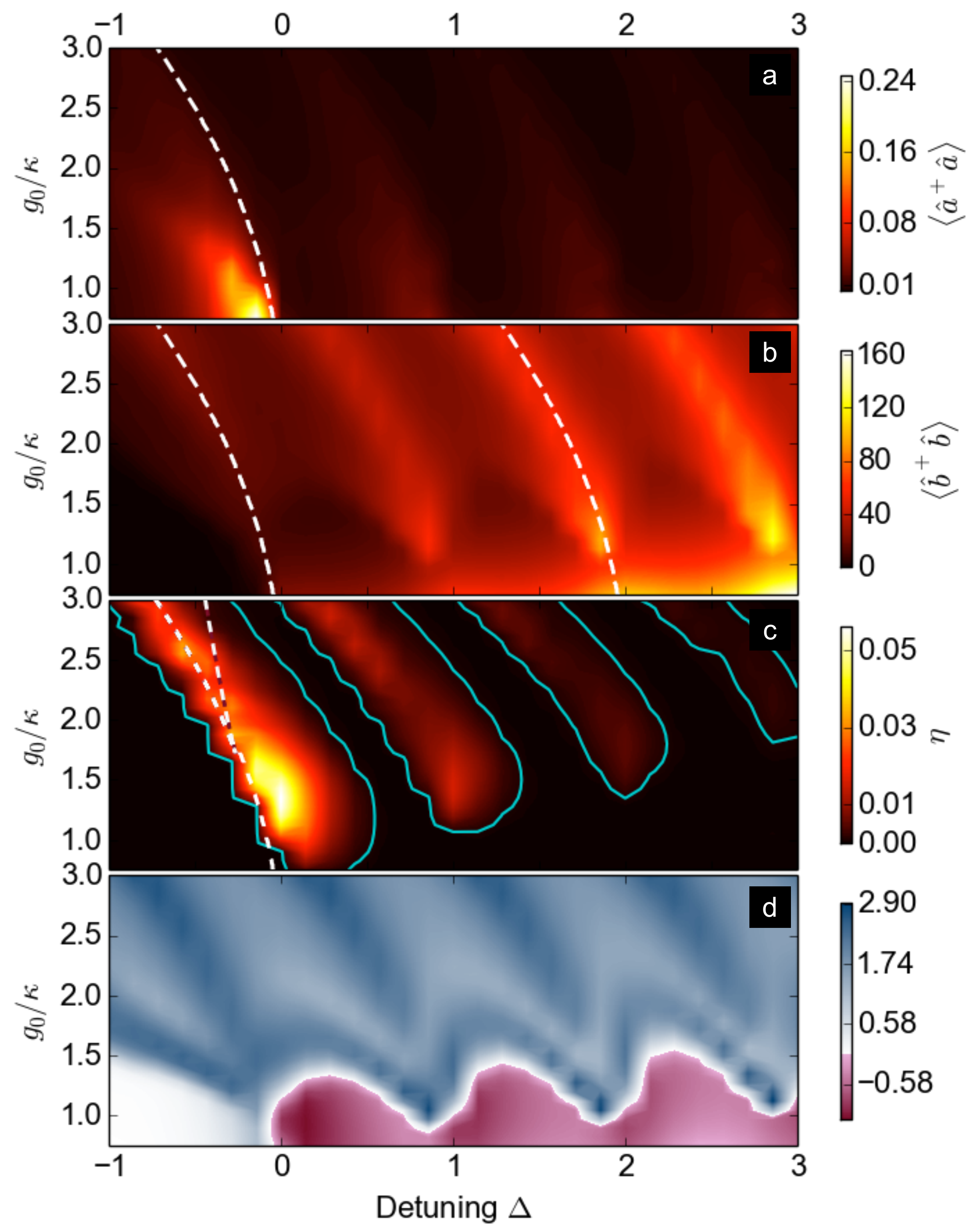}
\caption{(Color online) (a) Steady state cavity energy as a function of detuning and normalized coupling strength $g_{0}/\kappa$.   The dashed line shows the renormalized cavity frequency.  (b)  Steady state energy of the corresponding mechanical oscillator mode, including the first three oscillator sidebands.  The detuning value at which the sidebands occur follows the frequency-pulling of the cavity (dashed).  (c) Nonclassical ratio for the mechanical oscillator.  Contours (solid) show the regions where the nonclassical ratio is larger than $0.1\%$.  This region includes portions of the red-detuned side of the cavity response, below the renormalized cavity frequency, and the bistable region enclosed in the dashed lines. (d) Log-scale plot of the mechanical resonators Fano factor.}
\label{fig:fig2}
\end{figure}

Figure~\ref{fig:fig2} shows the results for a numerical simulation finding the steady-states of Eq.~(\ref{eq:master}), performed using QuTiP \cite{qutip1,*qutip2}, for a system with parameters $E=0.1, \kappa=0.3, Q_{m}=10^{4}$ and $\bar{n}_{\rm th}=0$, over a parameter space characterized by $-1\le \Delta \le 3$ and varying $g_{0}$ over the range $0.75\le g_{0}/\kappa \le 3$.  Note that the coupling strength is tunable in some superconducting circuit optomechanical realizations \cite{blencowe:2007}.  Here, our truncated Hilbert space includes four Fock states for the cavity mode and $200$ states for the mechanical resonator.  Simulation details are in the Appendix, and the source code can be obtained on the arXiv \cite{arxiv}.
\begin{figure}[t]
\begin{center}
\includegraphics[width=8.0cm]{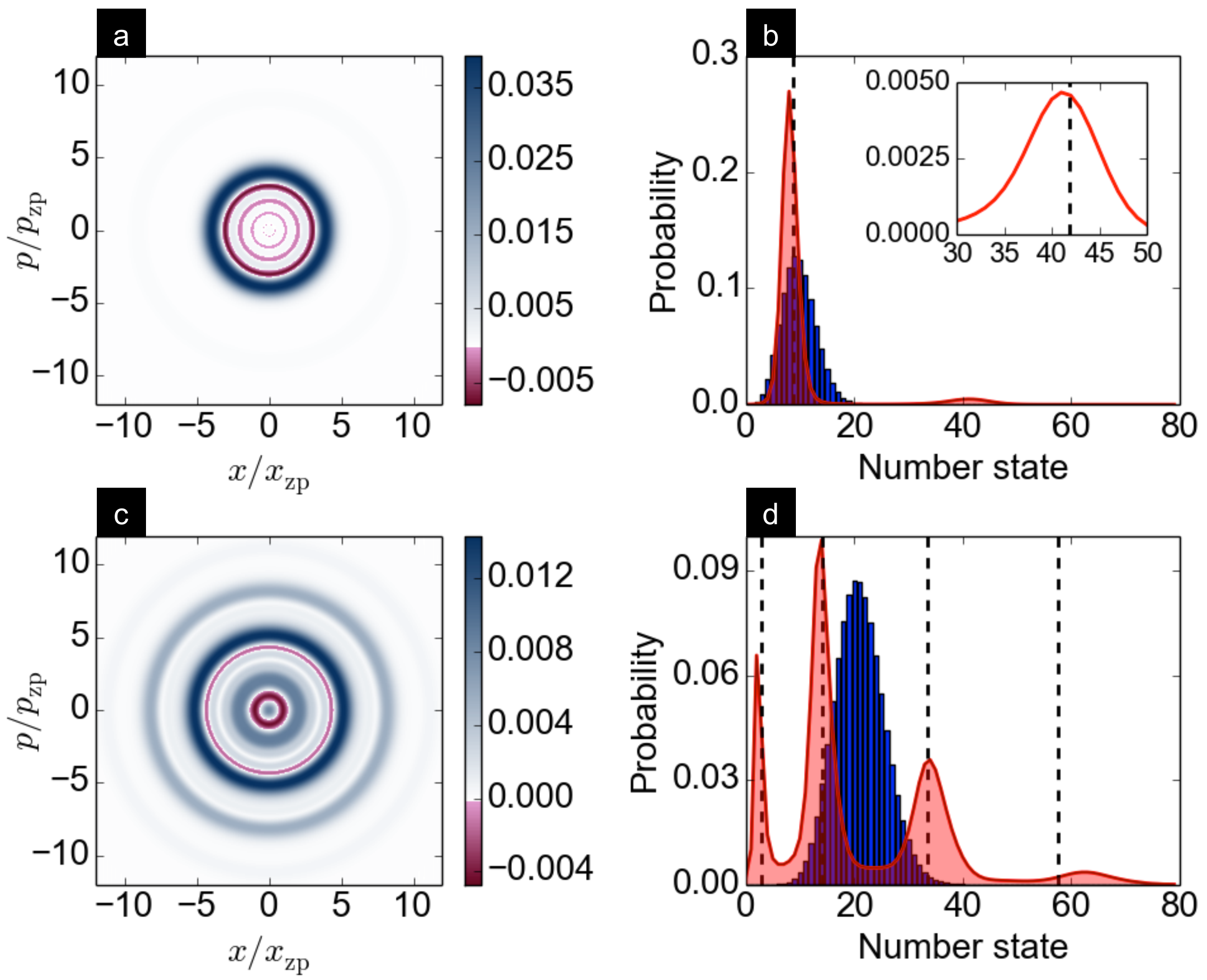}
\caption{(Color online) (a) Wigner distribution for the state at $\Delta=0, g_{0}/\kappa=1.35$ with the largest nonclassical ratio $\eta \simeq 6\%$.  (b) The number state probability distribution for the state in (a) (solid-red) together with a coherent state of the same amplitude (blue-bars).  Dashed lines show the corresponding semiclassical limit-cycle amplitudes found using Eqs.~(\ref{eq:force}) and (\ref{eq:power}).  Inset shows the small peak in the distribution for the larger limit-cycle that gives $F=5.2$. (c) Wigner function at $\Delta=-0.43, g_{0}/\kappa=2.4$, located in the bistable regime with $\eta\simeq 2\%$.  (d)  Phonon distribution function (solid-red) for state in (c) showing four stable limit-cycles with $F=10.4$, and an equal amplitude coherent state (blue-bars).}
\label{fig:fig3}
\end{center}
\end{figure}

First, as required for our analogue micromaser, Fig.~\ref{fig:fig2}a shows that the cavity photon number is well-below unity over the entire parameter range.  Second, we see in Fig.~\ref{fig:fig2}c that the strongest nonclassical states are generated when the cavity mode is driven just above its renormalized resonance frequency as determined by Eq.~(\ref{eq:nonlinear}), and at the mechanical sidebands.  These motional sidebands are not present in a conventional micromaser.  Surprisingly, this includes portions of the parameter space on the red-detuned side of the cavity, below the renormalized cavity resonance, and inside the bistable region of the cavity.  As discussed in detail below, the mechanical Fano factors given in Fig.~\ref{fig:fig2}d, $F=\langle(\Delta \hat{N}_{b})^{2}\rangle/\langle \hat{N}_{b}\rangle$, for the nonclassical states are typically larger than unity due to the presence of multiple stable oscillation amplitudes.  Furthermore, strong quantum features in the Wigner functions ($\eta\ge 1\%$) are not found at the higher resonator sidebands, as strong driving of the oscillator leads to multiple limit-cycles with overlapping number-state distributions that degrade the quantum signatures in these states.

The Wigner function and phonon probability distribution for two states in the nonclassical regions, the state with the largest nonclassical ratio, at $\Delta=0, g_{0}/\kappa=1.35$, and a state in the bistable region, at $\Delta=-0.43, g_{0}/\kappa=2.4$, are presented in Fig.~\ref{fig:fig3}.  The Wigner functions consist of an ensemble of rings, each corresponding to a stable limit-cycle, with a sub-Poissonian distribution.  Sub-Poissonian effects are well-known in both the micomaser \cite{filipowicz:1986} and Kerr-type nonlinear interactions \cite{buzek:1991}.  Here, the rotational symmetry arises because of phase diffusion \cite{rodrigues:2010}, and corresponds to a density matrix with only diagonal elements \cite{nunnenkamp:2011}.  Negative regions can be generated by single (Fig.~\ref{fig:fig3}a), or multiple (Fig.~\ref{fig:fig3}c) limit-cycles.  Although each limit-cycle is sub-Poissonian, the separation between oscillation amplitudes with nonzero occupation probabilities is responsible for the large Fano factors seen in Fig.~\ref{fig:fig2}d.  In general, the number of accessible limit-cycles increases with coupling strength and resonator occupation number.  Here, semiclassical limit-cycle energies can be directly equated to those for the quantum oscillator as static displacements of the oscillator are negligible in this regime.    

The onset, and subsequent decline, in the nonclassical ratio can be understood by fixing the detuning at $\Delta=0$, and sweeping the coupling from zero to $g_{0}/\kappa=3$.  This is comparable to varying the micromaser pump parameter, proportional to the coupling strength and atom-cavity interaction time \cite{walther:2006,filipowicz:1986}.  Here, the interaction time is the inverse of the cavity decay rate,  $\tau_{\rm{int}}=\kappa^{-1}$.  The resonator Q-factor also plays a role in the pump parameter, and should be sufficiently large to discern quantum effects.  To observe the switching between oscillator limit-cycles we pick the number state corresponding to the maximum probability amplitude in the density matrix as the order parameter \cite{rodrigues:2007}.
\begin{figure}[t]
\begin{center}
\includegraphics[width=8.0cm]{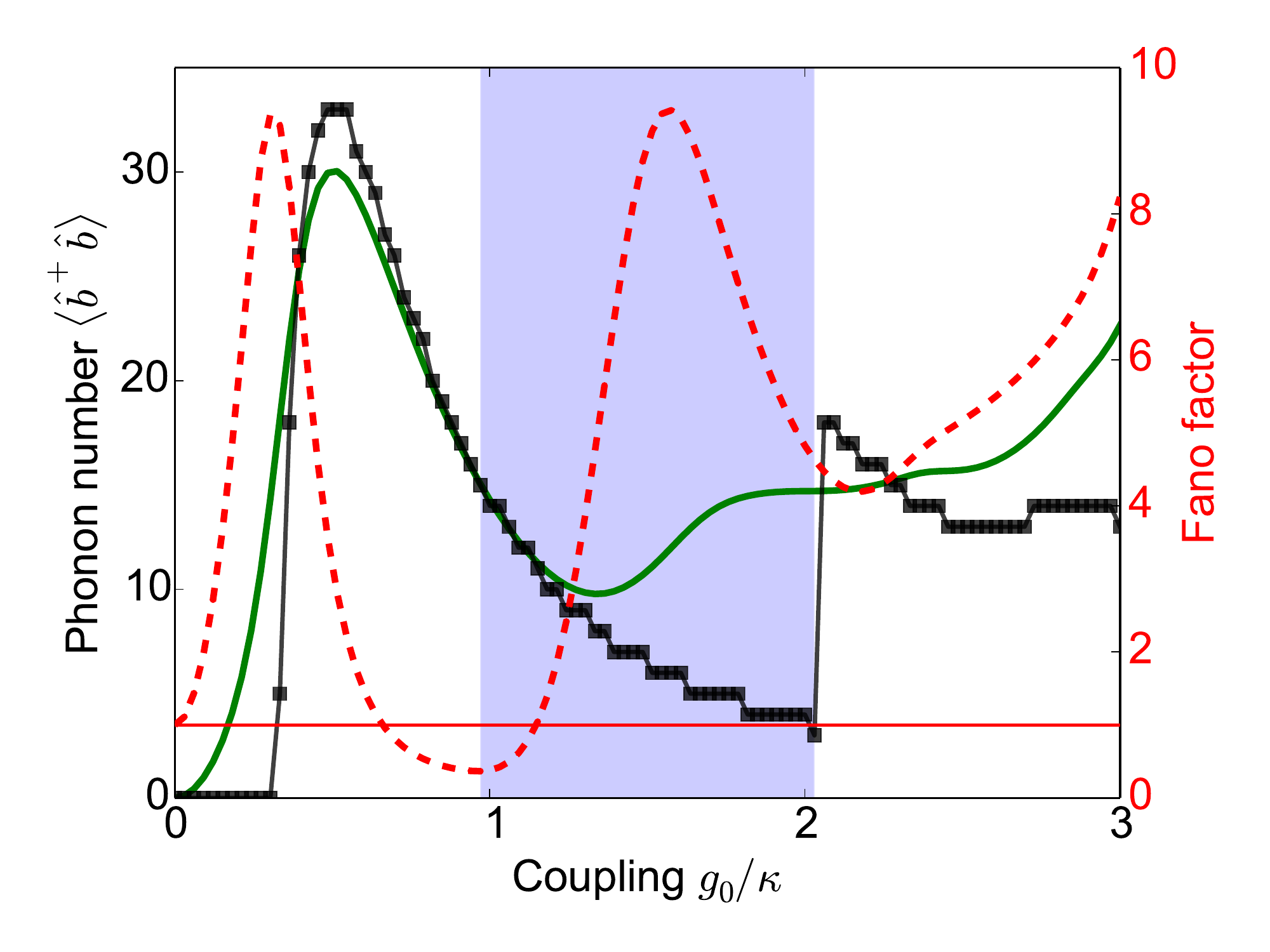}
\caption{(Color online) Oscillator order parameter (squares-black), mean phonon number (solid-green), and Fano factor (dashed-red) as a function of $g_{0}/\kappa$ at $\Delta=0$.  The shaded region indicates where $\eta\ge 1\%$.}
\label{fig:fig4}
\end{center}
\end{figure}

Figure~\ref{fig:fig4} shows order parameter transitions and Fano factors that are analogous to those seen in the micromaser \cite{filipowicz:1986}.  The initial transition between the $A=0$ fixed point and onset of limit-cycle oscillation indicates the value of $g_{0}$ at which the single-excitation interaction overcomes the intrinsic loss rate of the resonator, and corresponds to the micromaser threshold. Larger couplings give rise to sub-Poissonian statistics until another limit-cycle becomes accessible.  The subsequent discontinuous jump in the order parameter signals that this new limit cycle is now the preferred state of the resonator.  For the high-Q oscillator described here, this second transition is sharp even though we outside of the thermodynamic limit \cite{filipowicz:1986}, i.e. $\Gamma_{m}\rightarrow 0$.  Note that these features are fundamentally different than those observed in a laser or maser, where above threshold the cavity occupation number saturates, and the state of the cavity is a phase-randomized coherent state \cite{walls:2008}.  The transitions presented in Fig.~\ref{fig:fig4} are the characteristic signature of a micromaser, confirming that our optomechanical model is in fact an analogue of this system.

\begin{figure}[t]
\begin{center}
\includegraphics[width=8.5cm]{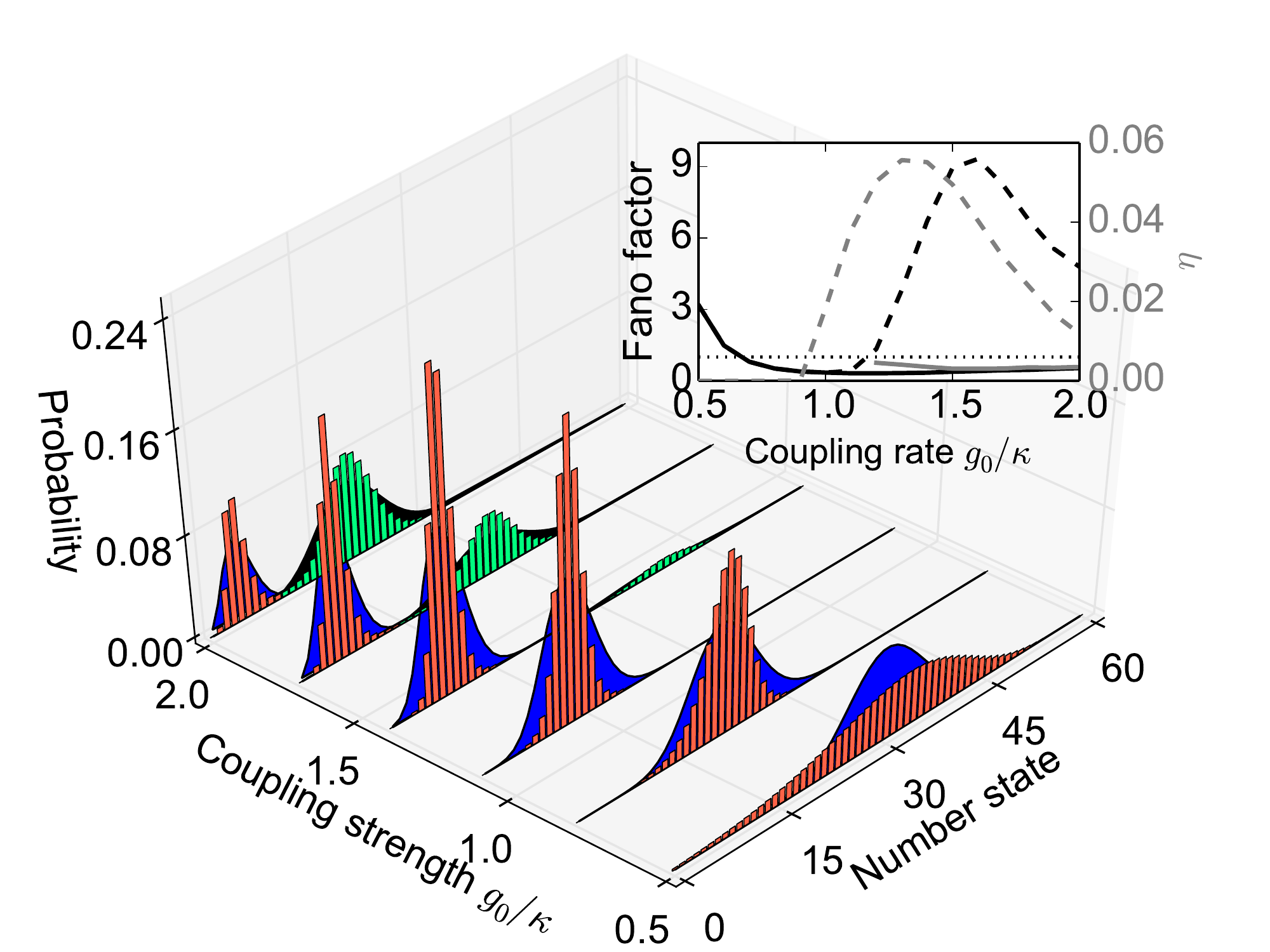}
\caption{(Color online) Distributions for the lower (red-bars) and upper (green-bars) limit-cycles along with corresponding coherent distributions, (blue-solid) and (black-solid), respectively of the same amplitude.  Coherent states are normalized to the probabilities of each limit-cycle.  Inset shows the Fano factors corresponding to the lower (black) and upper (grey) limit-cycles individually, as well as the Fano factor (dashed-black) and nonclassical ratio $\eta$ (dashed-grey) for the overall state.  Above $g_{0}/\kappa \simeq 1.3$, the two limit-cycles begin to merge, and the individual limit-cycle Fano factors are approximate values.}
\label{fig:fig5}
\end{center}
\end{figure}

Figure~\ref{fig:fig5} shows several resonator distributions along with effective Fano factors corresponding to the individual limit-cycles responsible for the oscillations in Fig.~\ref{fig:fig4}, and shown in Fig.~\ref{fig:fig3}b.  Both limit-cycles are clearly sub-Poissonian when $\eta>0$, although the state itself can have a large Fano factor. At $g_{0}/\kappa\sim 1.3$, the limit-cycle distributions begin to overlap and the description in terms of individual oscillation amplitudes is no longer valid; the overall distribution of the resonator, which is super-Poissonian, determines the nonclassical features.  This merger causes a marked reduction in the nonclassical features of the oscillator Wigner functions.  This effect is pronounced at the oscillator sidebands where large phonon numbers inherently give rise to multiple overlapping limit-cycles.

\begin{figure}[t]
\begin{center}
\includegraphics[width=8.0cm]{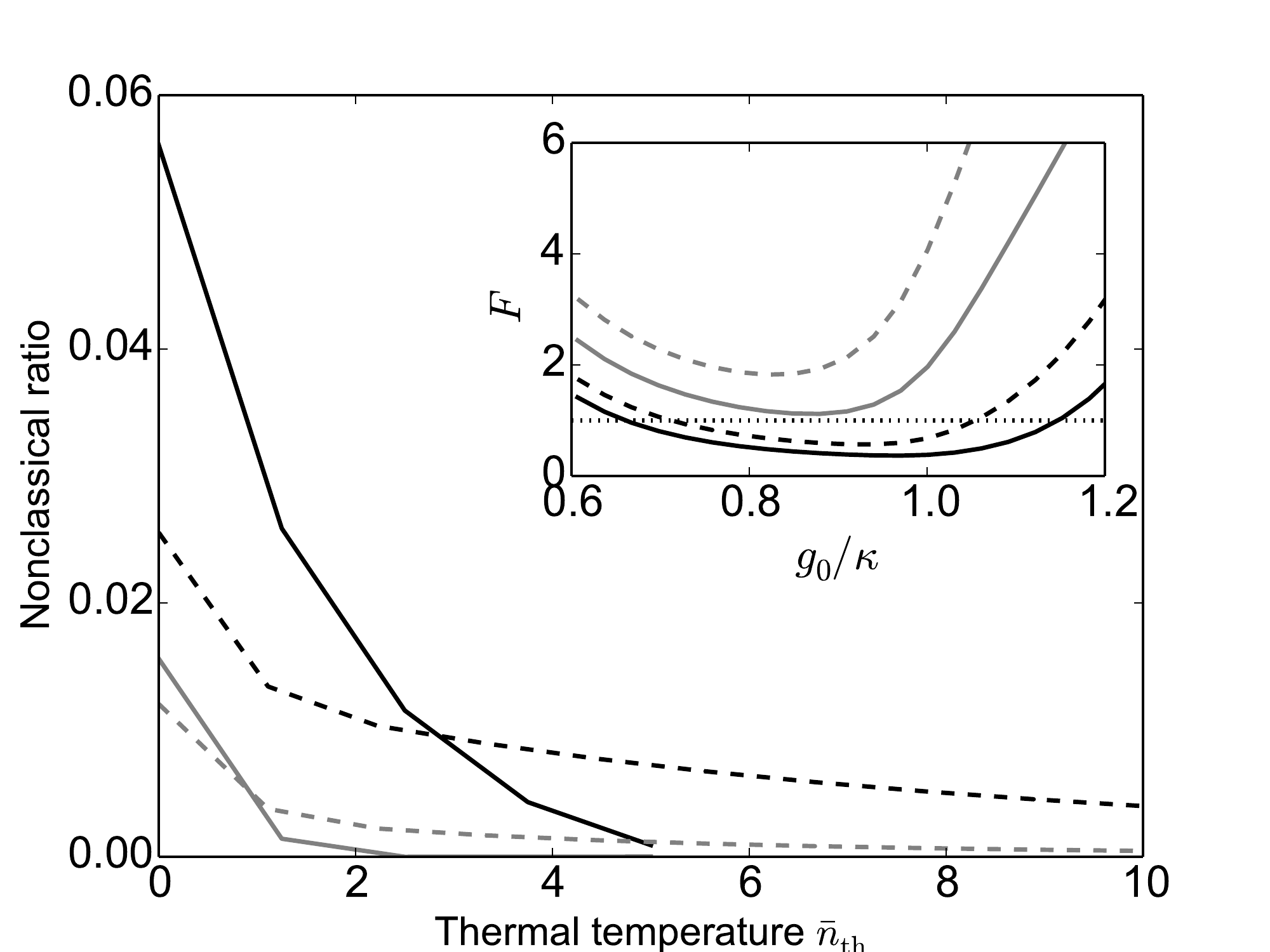}
\caption{Nonclassical ratio for the states in Fig.~\ref{fig:fig3}a (black) and Fig.~\ref{fig:fig3}c (dashed-black), as well as two states on the first mechanical sideband, $\Delta=1, g_{0}/\kappa=1.4$ (grey) and $\Delta=0.5, g_{0}/\kappa=2.6$ (dashed-grey), as a function of the bath temperature $\bar{n}_{\rm th}$.  Inset shows the Fano factors for the $\Delta=0$ limit-cycle above threshold seen in part (a) for bath temperatures $\bar{n}_{\rm th}=0$ (black), $1$ (dashed-black), $3$ (grey), and $5$ (dashed-grey).}
\label{fig:fig6}
\end{center}
\end{figure}
Finally, as in the micromaser \cite{davidovich:1996}, the addition of a nonzero thermal environment for the oscillator quickly masks the quantum features in our optomechanical analogue.  In Fig.~\ref{fig:fig6} we highlight this decay, via Eq.~(\ref{eq:master}), as a function of the bath occupation number $\bar{n}_{\rm th}$ for both the nonclassical ratio taken at several points in parameter space, and Fano factors for the initial limit cycle at $\Delta=0$.  It is seen that the introduction of even a low temperature thermal environment significantly diminishes the quantum features of the mechanical states.  However, at larger coupling strengths, some residual quantum features do persist at higher temperatures.

\section{Conclusion}\label{sec:conclusion}
We have shown that nonclassical states of a mechanical oscillator can be generated in an optomechanical analogue of the micromaser when the cavity is damped so as to be occupied by at most a single photon. This system exhibits strong sub-Poissonian limit-cycles, nonclassical Wigner functions, and phonon oscillations in the resonator that are the signature features of a micromaser.  These features are reduced at nonzero bath temperatures as the increased fluctuations interfere with the coherent cavity-resonator interaction.  Note that trapped states can not be produced in this setup as these rely on the transient two-level atoms undergoing an integer number of Rabi oscillations \cite{weidinger:1999}.  In addition, the micromaser analogy suggests that the presence of multiple photons simultaneously in the cavity mode should degrade the nonclassical signatures of the resonator \cite{walther:2006}; the quantum properties of the oscillator should vanish in the strong-driving, or high-Q cavity limits.  This also follows from standard optomechanical theory where it is well-known that Eq.~(\ref{eq:hamiltonian}) is effectively linearized in the strong driving limit \cite{aspelmeyer:2013}.  However, understanding where the crossover occurs requires a more complete analytic theory, or more robust numerical methods capable of analyzing multiple oscillator modes with large occupation numbers.  The work presented here is, to the best of our knowledge, the first time that these micromaser characteristics have been predicted in a system with no atom-like component, and this link helps to further our understanding on the generation of quantum states in macroscopic mechanical resonators.

Note that, during the submission process, we became aware of Ref.~\cite{lorch:2013}  that derives, via laser theory, analytic expressions for the case of a single limit-cycle of a high-Q oscillator in the regime where $g_{0}/\kappa \lesssim 1$.  Importantly, this work indicates that nonclassical states can be generated outside of the single-excitation regime.  However, these states show markedly less non-classicality than the corresponding states in the single-photon regime.  A result that is inline with the micromaser theory presented here that indicates that multiple cavity photons will diminish nonclassical features of the resonator.

\appendix
\section{Numerical Methods}\label{sec:app}
Simulations of Eq.~(\ref{eq:master}) were carried out using a truncated Hilbert space consisting of the four lowest Fock states for the cavity and 200 states for the mechanical resonator.  Internally, QuTiP solves for the steady state density matrix of the combined cavity+resonator system using a sparse LU decomposition method to evaluate
\begin{equation}
\left(\mathcal{L}+M\right)\vec{\rho}=\left(\begin{array}{c}1 \\0 \\\vdots\end{array}\right); \ \ M\vec{\rho}=\left(\begin{array}{c}\rm Tr(\rho) \\0 \\\vdots\end{array}\right),
\end{equation}
where we have used the unit-trace property of the density matrix.  For the number of states in the Hilbert space, this is equivalent to finding the inverse of the Liouvillian super-operator with $\sim 4\times 10^{11}$ elements.  To map out the entire parameter space, we divide the detuning and coupling strength axes into $29$ and $31$ partitions, respectively, for a total of $899$ elements and solve each for the steady state density matrix.  Evaluating the commutator for both modes allows for checking whether the number of states in our truncated model is in fact adequate.  We have verified that the values for both the cavity and oscillator commutators are near one over the entire parameter space suggesting that we have included an appropriate number of basis states in our simulations.

Calculations of the Wigner function for the mechanical density matrix were performed using a Fourier transform based method as more standard numerical techniques, based on Laguerre polynomials for example, can not calculate the Wigner function for states with expectation values $\langle N\rangle \gtrsim 50$ when using standard double precision arithmetic.  For each density matrix, the Wigner function was evaluated at $\sim2.5\times 10^{6}$ points over the interval $[-25,25]$ in units of the oscillators zero-point amplitude.  The resulting area element $dxdp$ is much smaller than the typical feature size present in the Wigner functions, and therefore inadvertent averaging will not affect the calculation of the nonclassical ratio.  This Wigner area is also large enough to take into account all of the limit-cycles presented in the text.

\section*{Acknowledgements}
I would like to thank A. D. Armour, M.-S. Choi, J. R. Johansson and H. Kang for helpful discussions.  This work was supported by startup funding from Korea University.

\bibliography{refs}
\end{document}